# Understanding Epileptiform After-Discharges as Rhythmic Oscillatory Transients

Gerold Baier[1], Peter N. Taylor[2, 3, 4*] and Yujiang Wang[2, 3, 4*]

[1] Cell and Developmental Biology, University College London, London, UK, [2] Institute of Neuroscience, Newcastle University, Newcastle upon Tyne, UK, [3] Interdisciplinary Computing and Complex BioSystems (ICOS), School of Computing Science, Newcastle University, Newcastle, UK, [4] Institute of Neurology, University College London, London, UK

Electro-cortical activity in patients with epilepsy may show abnormal rhythmic transients in response to stimulation. Even when using the same stimulation parameters in the same patient, wide variability in the duration of transient response has been reported. These transients have long been considered important for the mapping of the excitability levels in the epileptic brain but their dynamic mechanism is still not well understood. To investigate the occurrence of abnormal transients dynamically, we use a thalamo-cortical neural population model of epileptic spike-wave activity and study the interaction between slow and fast subsystems. In a reduced version of the thalamo-cortical model, slow wave oscillations arise from a fold of cycles (FoC) bifurcation. This marks the onset of a region of bistability between a high amplitude oscillatory rhythm and the background state. In vicinity of the bistability in parameter space, the model has excitable dynamics, showing prolonged rhythmic transients in response to suprathreshold pulse stimulation. We analyse the state space geometry of the bistable and excitable states, and find that the rhythmic transient arises when the impending FoC bifurcation deforms the state space and creates an area of locally reduced attraction to the fixed point. This area essentially allows trajectories to dwell there before escaping to the stable steady state, thus creating rhythmic transients. In the full thalamo-cortical model, we find a similar FoC bifurcation structure. Based on the analysis, we propose an explanation of why stimulation induced epileptiform activity may vary between trials, and predict how the variability could be related to ongoing oscillatory background activity. We compare our dynamic mechanism with other mechanisms (such as a slow parameter change) to generate excitable transients, and we discuss the proposed excitability mechanism in the context of stimulation responses in the epileptic cortex.

Keywords: afterdischarge, epilepsy model, spike-wave, stimulation, transients

## 1. INTRODUCTION

Epileptic seizures are typically marked by abnormal rhythmic discharges of electrical activity in the human brain. The rhythmicity is taken as an indication of an underlying deterministic nonlinear oscillation. The transition from a disorganized background activity to the epileptic rhythm is described as a state transition either due to a changing parameter (Breakspear et al., 2006); or due to a cross-separatrix perturbation in a bistable situation (Lopes da Silva et al., 2003); or a combination of both (Wang et al., 2014). A problem that has not been satisfactorily addressed is the nature of the termination of the abnormal rhythm. Corresponding to the situation at rhythm onset, rhythm termination could be equally be due to a parameter change (Kramer et al., 2012), a cross-separatrix perturbation in a bistable situation (Taylor et al., 2014) or a mixture of the two. However, in the case of human EEG, no direct observation for either the claimed (slow) parameter change or the cross-separatrix perturbation has been provided, although we discuss some indirect evidence later on.

Abnormal rhythmic discharges can also often be induced in patients with epilepsy by stimulation. Well-known examples are the reflex epilepsies, where epileptic rhythms are induced by some form of motor-sensory stimulation (Koepp et al., 2016), and the so-called afterdischarges following invasive electrical stimulation of the cortex (Penfield and Jasper, 1954; Blume et al., 2004). **Figure 1** is an example of an abnormal spike-wave rhythm induced by transcranial magnetic stimulation. As in the case of spontaneous epileptic seizures, these rhythms typically self-terminate and the dynamic mechanism of the termination remains speculative.

A previous computational modeling study has suggested that there is a third possibility to explain the termination of an abnormal rhythm. Looking at the case of electrical responses to cortical pulse stimulation, it was suggested that prolonged rhythmic transients could be self-terminating in spatially extended systems (Goodfellow et al., 2012a; Taylor et al., 2013). This means that neither a parameter change nor a cross-separatrix stimulus would be required for the rhythm to stop. The transient abnormal rhythm would then constitute an example of excitability of a dynamical system. Excitability is a prominent feature of information-processing systems (like neurons) as it allows self-resetting after information has been delivered. It is unknown whether the epileptic brain constitutes an excitable medium with thresholds to epileptic discharges, but computational models demonstrate that a large number of features surrounding epilepsy can be recreated in heterogeneous excitable media (Goodfellow et al., 2012b). However, a major problem with high-dimensional spatio-temporal models is that the mathematical nature of state space features is difficult to unveil.

Here we study a previously proposed model of cortico-thalamic interaction which ignores the spatial extension and heterogeneities of the cortex as is typically done to describe widely synchronized epileptiform rhythms shown in e.g., Blume et al. (2004) and Kimiskidis et al. (2015), also see **Figure 1** for an example. To investigate the dynamic mechanisms of the transients in the thalamo-cortical model, we propose a simplified version of it to demonstrate the possibility of abnormal rhythmic transients in its low-dimensional state space. With some variations, different types of transient waveforms can be created in the full model, in particular those known to occur in patients with epilepsy. With this proposed alternative mechanism, we can explain the observation of variability on stimulation response in the clinical setting, and suggest a model based prediction to explain the variability. Future computational studies of abnormal brain activity can use these mechanisms to generate alternative predictions to be compared to clinical observations (instead of attempting to fit a single model to the data), c.f. Estrada et al. (2016).

## 2. METHODS
### 2.1. Thalamocortical Model

We simulate thalamocortical interactions by following previous modeling based on the known connectivity of this system (see **Figure 2B**; c.f. Suffczynski et al., 2004; Breakspear et al., 2006). Specifically, the neural mass approach by Suffczynski et al. (2004) forms a neural population version of the detailed biophysical model described by Destexhe (1998). In our macroscopic model, we consider the activity changes in four neural populations. The cortical pyramidal cell population (*PY*) is self-excitatory (Amari, 1977) and excites the cortical inhibitory interneuron population (*IN*) (Suffczynski et al., 2004). In addition, *PY* excites the thalamocortical cell population in the thalamus (*TC*), and a population of cells in the reticular nucleus of the thalamus (*RE*) (Suffczynski et al., 2004; Yousif and Denham, 2005). The interneuron population *IN* inhibits the excitatory cortical *PY* population (Suffczynski et al., 2004). Direct thalamic output to the cortex comes exclusively from excitatory *TC* connections to the *PY* population (Breakspear et al., 2006). Intrathalamic connectivity is incorporated into the model as follows: *TC* cells have excitatory projections to *RE*, which in turn inhibits the *TC* population along with self-inhibition of *RE*. This connectivity scheme is consistent with experimental results reviewed in Pinault and O'Brien (2005) and summarized in their Figure 1.

It was demonstrated in a minimal model of epileptic spike-wave discharges (SWD) that at least one slow driver is required in addition to the cortical *PY* and *IN* units (Wang et al., 2012).

There is experimental evidence for abnormal slow thalamic processes (variations in a tonic inhibitory current), which may be a common mechanism in typical absence seizures (Cope et al., 2009). This is also supported by theoretical studies that find slow timescales crucial for the generation of realistic SWD. These studies either incorporate the slow timescale directly by modeling the slower reaction of thalamic populations (Marten et al., 2009; Taylor et al., 2013) or by incorporating explicit delays (Breakspear et al., 2006). Marten et al. (2009) compares these approaches and finds bifurcation structures leading to the onset of SWD which are similar. As the exact dynamic mechanisms underlying the emergence of the slow timescale is still unclear, we assume here that the thalamic compartment operates on a slower timescale in the abnormal condition of increased susceptibility to undergo transitions to SWD. From a theoretical point of view, this has the advantage that the model can be analyzed in terms of slow-fast subsystems (Wang et al., 2012).

The model uses the neural population approach based on the neural field equations proposed by Amari (1977). Spike-wave solutions are found following the analysis done in Taylor

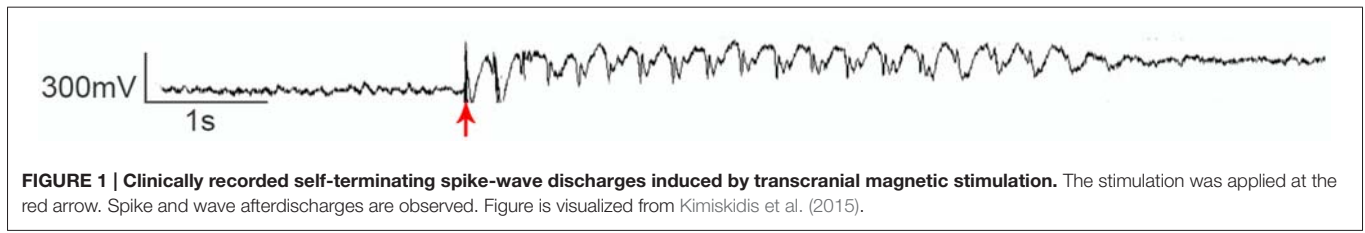

**FIGURE 1 | Clinically recorded self-terminating spike-wave discharges induced by transcranial magnetic stimulation.** The stimulation was applied at the red arrow. Spike and wave afterdischarges are observed. Figure is visualized from Kimiskidis et al. (2015).

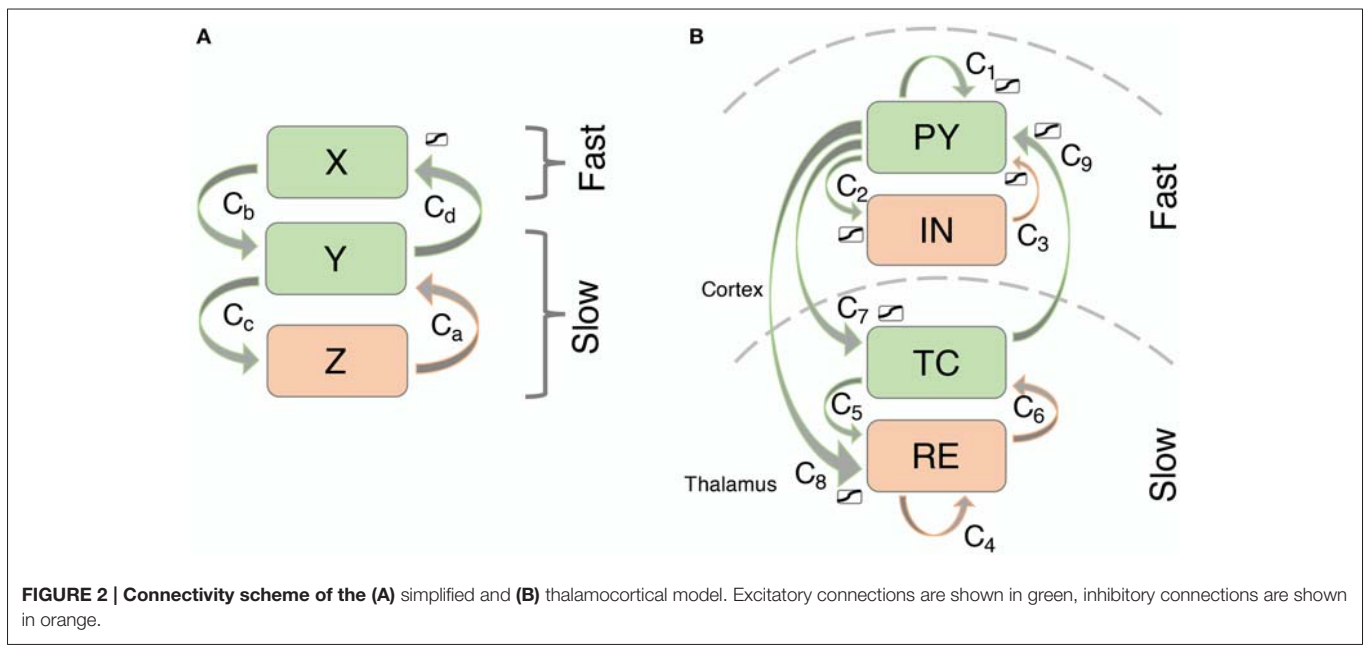

**FIGURE 2 | Connectivity scheme of the (A)** simplified and **(B)** thalamocortical model. Excitatory connections are shown in green, inhibitory connections are shown in orange.

and Baier (2011) and Taylor et al. (2013). The set of ordinary differential equations is given as:

$$\frac{dPY}{dt} = \tau_1(h_{py} - PY + C_1 f[PY] - C_3 f[IN] + C_9 f[TC])$$
$$\frac{dIN}{dt} = \tau_2(h_{in} - IN + C_2 f[PY])$$
$$\frac{dTC}{dt} = \tau_3(h_{tc} - TC + C_7 f[PY] - C_6 (s[RE]))$$
$$\frac{dRE}{dt} = \tau_4(h_{re} - RE + C_8 f[PY] - C_4 (s[RE]) + C_5 (s[TC]))$$
(1)

where $h_{py,in,tc,re}$ are input parameters, $\tau_{1...4}$ are time scale parameters and $f[.]$ and $s[.]$ are the activation functions:

$$f[u] = (1/(1 + \epsilon^{-u}))$$ (2)
$$s[u] = au + b$$ (3)

with $u = PY, IN, TC, RE$. The parameter $\epsilon$ determines the slope of the sigmoid. Two activation functions are used here, one sigmoid, one linear. This is because we found that in our previous studies (Taylor et al., 2014) the $RE$ and $TC$ interactions mainly happened in the linear range of the sigmoid activation function.

Hence we replaced them with a linear activation. This is also useful to directly compare the slow thalamic subsystem to the simplified model later, where we also use linear activation for simplification.

Details about the parameters can be found in **Tables S1**, **S2**. Code to simulate the model can be found online at https://senselab.med.yale.edu/modeldb/showmodel.cshtml?model=168856.

## 2.2. Simplified Model

To investigate mechanisms leading to bistability/excitability in the model we use a simplified model of three variables which allows for easier visualization (we will call it the 3V model - **Figure 2A**). The simplified model preserves the important aspects of the more detailed model including timescale separation and bistability. Specifically, the two slow variables create a focus in the slow subsystem, just like the full thalamo-cortical model. The two fast cortical variables have been simplified to one, effectively removing the Hopf bifurcation in the fast subsystem. This is so that the full system is simplified to only display simple oscillations, rather than slow-fast oscillations. Indeed, the transient dynamics are built up by these two slow and one fast variables. As we will show in results, the additional fast variable in the full system just adds the Hopf bifurcation back into the system, allowing for transient slow-fast (spike-wave) oscillations. The equations for the simplified system are as follows:

$$\frac{dX}{dt} = \tau_{fast}(h_X - X + C_d f[Y])$$
$$\frac{dY}{dt} = \tau_{slow}(h_Y - Y - C_a Z + C_b X)$$
$$\frac{dZ}{dt} = \tau_{slow}(h_Z - Z + C_c Y) \quad (4)$$

where $h_{X,Y,Z}$ are input parameters, $\tau_{slow,fast}$ are time scale parameters and $f[.]$ is the activation function above (Equation 2).

We can essentially identify the slow subsystem (*TC* and *RE*) with the slow subsystem in the 3V model (*Y* and *Z*). The cortical system *PY* and *IN* can be considered the expansion of the 3V system variable *X*. The parameter values for the simplified model were obtained through visual comparison of the state space structures as described above. Or in other words, we constructed the state space structures in the simplified model with the parameters, such that they matched the the full system as described above.

Details about the parameters can be found in **Tables S1, S2**.

## 2.3. Slow and Fast Manifolds

We will use the slow and fast manifolds in state space to explain some of the phenomena and mechanisms we observe. The slow manifold is technically defined as the manifold in state space on which the change in the fast subsystem is zero, i.e., the slow dynamics dominate. For the 3V model, this would be the condition $\frac{dX}{dt} = 0$, hence $(h_X - X + C_d f[Y]) = 0$, $h_X + C_d f[Y] = X$. This means the manifold is a sigmoid-shaped plane (green plane in **Figure 3**). For the full thalamo-cortical system, this would be the conditions $\frac{dPY}{dt} = 0, \frac{dIN}{dt} = 0$. Essentially the fixed point of the fast subsystem. However, in our analysis, we have expanded this definition of slow manifold, by also including limit cycles of the fast subsystem in the slow manifold (blue mesh in the later **Figures 6–8**). This is justified in more detail in Wang et al. (2012), but essentially we assume that the time scale separation is sufficient to separate the slow dynamics from the fast oscillations. We obtained the slow manifold for the full system numerically by obtaining the fixed points and limit cycles of the fast subsystem. The numerical simulation approach was used here (as opposed to continuation), as routine for detecting fixed points and limit cycles could be reused for the detection of transient events with small adaptations. In both cases we essentially detect if trajectories stay in the proximity of each other.

The fast manifold, in contrast, is where the slow dynamics are essentially zero. For the 3V system this means $\frac{dY}{dt} = 0$ and $\frac{dZ}{dt} = 0$. This gives $(h_Y - Y - C_a Z + C_b X) = 0$ and $(h_Z - Z + C_c Y) = 0$. If expressed as dependent on X we get: $Z = h_Z + C_c Y$ and $Y = (-C_a * h_Z + C_b * X + h_Y)/(1 + C_a * C_c)$. In other words, the fast manifold is a line in the state space (orange line in **Figure 3**). For the full thalamocortical system, we have the conditions: $\frac{dTC}{dt} = 0, \frac{dRE}{dt} = 0$. This gives a linear equation system, where the solution is the set of fixed points of the slow subsystem (orange line in **Figure 6F**).

## 3. RESULTS

### 3.1. Analysis of a Slow-Fast Subsystem

To begin our analysis, we will look at the 3V simplified system to illustrate the main dynamic mechanisms before applying the analogous analysis to the full thalamo-cortical model.

The reduced three-dimensional model is a slow-fast system, with one fast variable (*X*) and two slow variables (*Y*, *Z*). Previous studies of such slow-fast systems have shown that Fenichels Theorem (Fenichel, 1979) is a powerful tool of analysis. The theorem essentially states that the (stable of unstable) fixed points of the fast subsystem form a manifolds in the full state space with similar properties. By analysing these manifolds, one can understand the dynamics of the full system in state space.

Using Fenichels Theorem, and the slow and fast manifolds, the behavior of the system can be understood geometrically (Izhikevich, 2000; Desroches et al., 2012; Wang et al., 2012). These slow and fast manifolds are shown in **Figures 3A–D** as green and red manifolds, respectively. The slow manifold (i.e., where the change in the fast subsystem is zero, $dX/dt = 0$) is a green sigmoid shaped plane, which is attractive in the direction of *X*. The fast manifold (i.e., where there is no change in the slow subsystem, $dY/dt = 0$ and $dZ/dt = 0$) is a red line indicating a stable focus in the slow subsystem. The intersection of these two manifolds is a fixed point of the full system (blue dot in **Figures 3A–D**). Additionally, when the fast manifold (red) is positioned near the non-linear part of the slow sigmoidal manifold (green), the interplay between the fast and slow dynamics can create a stable limit cycle in the full system (indicated in magenta in **Figures 3A–D**) that is bistable to the fixed point (blue). The bistability is additionally demonstrated in **Figures 3E,F** by perturbing the fixed point and showing the following trajectory evolving to the limit cycle; and by a subsequent perturbation from the limit cycle back to the fixed point. Note that the bistable limit cycle is created around (and due) to the non-linearity in the sigmoidal slow manifold.

The stable fixed point and the stable limit cycle are separated by a saddle cycle (indicated in orange in **Figures 3A–D**). We additionally show the stable direction of the saddle cycle as a gray tube manifold in **Figures 3A–D**. This manifold also represents the separatrix between the fixed point and limit cycle.

In order to understand the type of bifurcation leading to this bistability, we slowly vary the parameter $h_X$, which controls how close the fast manifold (red line) is shifted toward the nonlinear part of the sigmoid-shaped slow manifold (green plane). We find that the new coexisting limit cycle arises in a fold of cycles bifurcation (marked as FoC1 in **Figure 4A**), as the fast manifold approaches the nonlinear part of the sigmoidal slow manifold. When $h_X$ is further increased, the stable fixed point becomes unstable in a subcritical Andronov-Hopf bifurcation, where also the saddle cycle disappears (marked as H1 in **Figure 4A**). Due to the symmetry in the sigmoid slow manifold, both bifurcations occur again as the fast manifold passes through the upper part of the sigmoid (marked as H2 and FoC2 in **Figure 4A**).

Interestingly, the impending fold of cycles bifurcation introduces a slowing/trapping in the local state space, which is well documented for the fold or saddle-node types of bifurcations (see e.g., Damme and Valkering, 1987). In the case of a fold of cycles bifurcation, this means that long transients of oscillations can be observed prior to the bifurcation. **Figure 4B** shows an example of such a long transient lasting for 14 cycles (over 5 model time seconds). **Figure 4C** shows the same time series in state space. In this projection the slowing/trapping in state space becomes particularly clear, as the trajectories dwell in the region in state space, where the stable and saddle limit cycle are to be born.

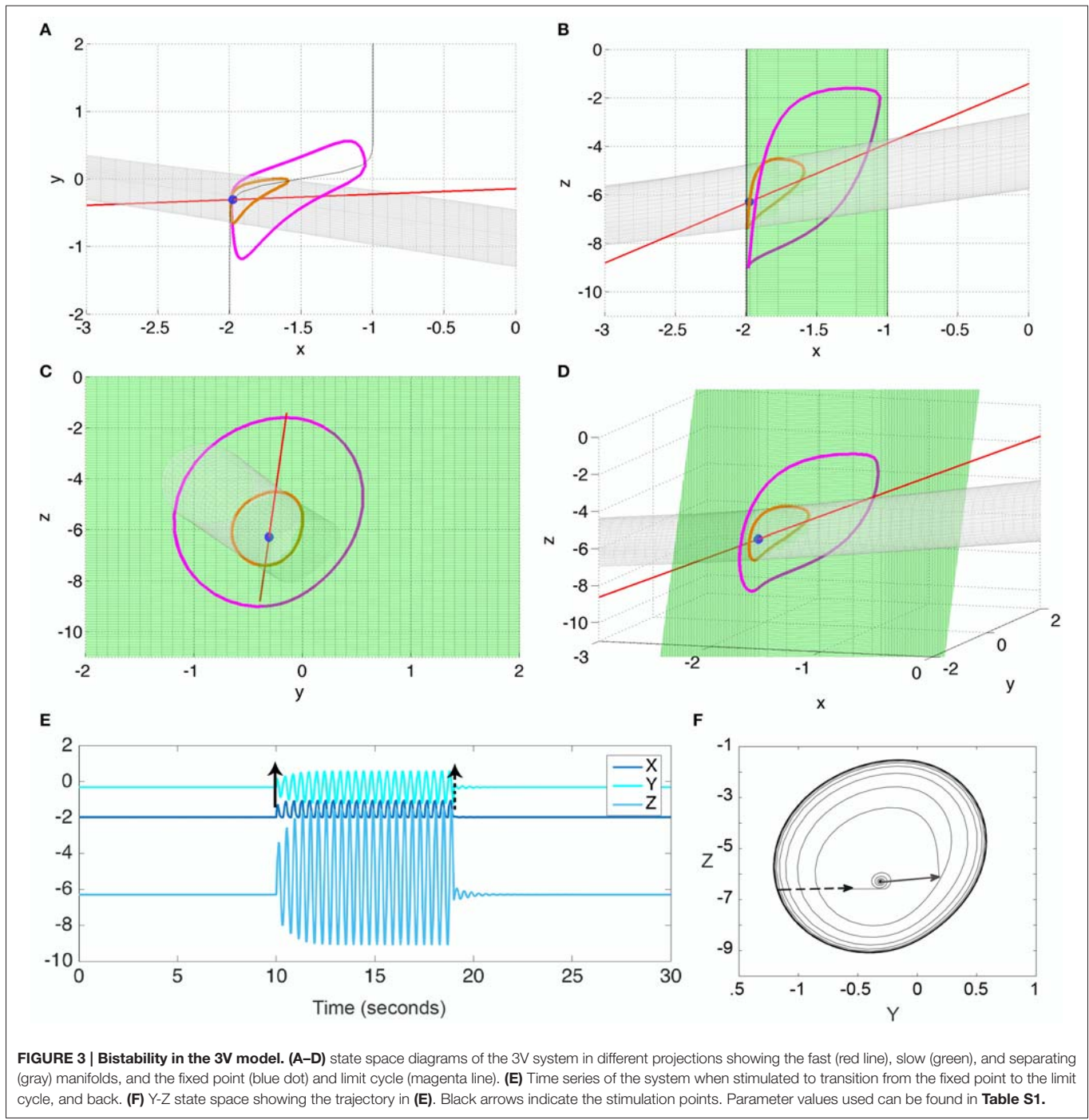

FIGURE 3 | Bistability in the 3V model. (A–D) state space diagrams of the 3V system in different projections showing the fast (red line), slow (green), and separating (gray) manifolds, and the fixed point (blue dot) and limit cycle (magenta line). (E) Time series of the system when stimulated to transition from the fixed point to the limit cycle, and back. (F) Y-Z state space showing the trajectory in (E). Black arrows indicate the stimulation points. Parameter values used can be found in **Table S1**.

We additionally measured the length of these transients for different values of $h_X$ and find a sudden increase just prior to the FoC bifurcation in parameter space, (**Figure 5**, first column). This behavior changes quantitatively with other parameters. For example, as the bifurcation point in $h_X$ shifts with the parameter $\tau_{slow}$, the parameter region in $h_X$ supporting long transients also increases. In other words, long transients can be found further away from the FoC point in $h_X$ as $\tau_{slow}$ decreases (**Figure 5A**). When analysing properties of the transients further, we observe that indeed the oscillation appears to become slower with decreasing $\tau_{slow}$ (**Figure 5B**). This results in a roughly constant number of oscillations for different $\tau_{slow}$ (**Figure 5C**). However, this effect is not observed in another parameter (e.g., $C_c$, **Figures 5D–F**). The parameter region in $h_X$ showing long transients appears to stay about the same for different $C_c$ (**Figure 5D**), but the number of oscillations increases substantially with different $C_c$ (**Figure 5F**). These two examples illustrate that the properties of the transients near the FoC bifurcations in $h_X$ can be modulated by other parameters. Particularly, the properties of the duration of the transient and the number of oscillations in the transient can be modified. We found that the modification essentially works by the mechanism of changing the flow in state space near the ghost of the saddle cycle. More precisely, the local flow creating the dwelling behavior in state space (near where the saddle node of cycles bifurcation will happen) determines how the transients will appear. This local flow is influenced by other parameters of the system, such as $C_c$ and $\tau_{slow}$.

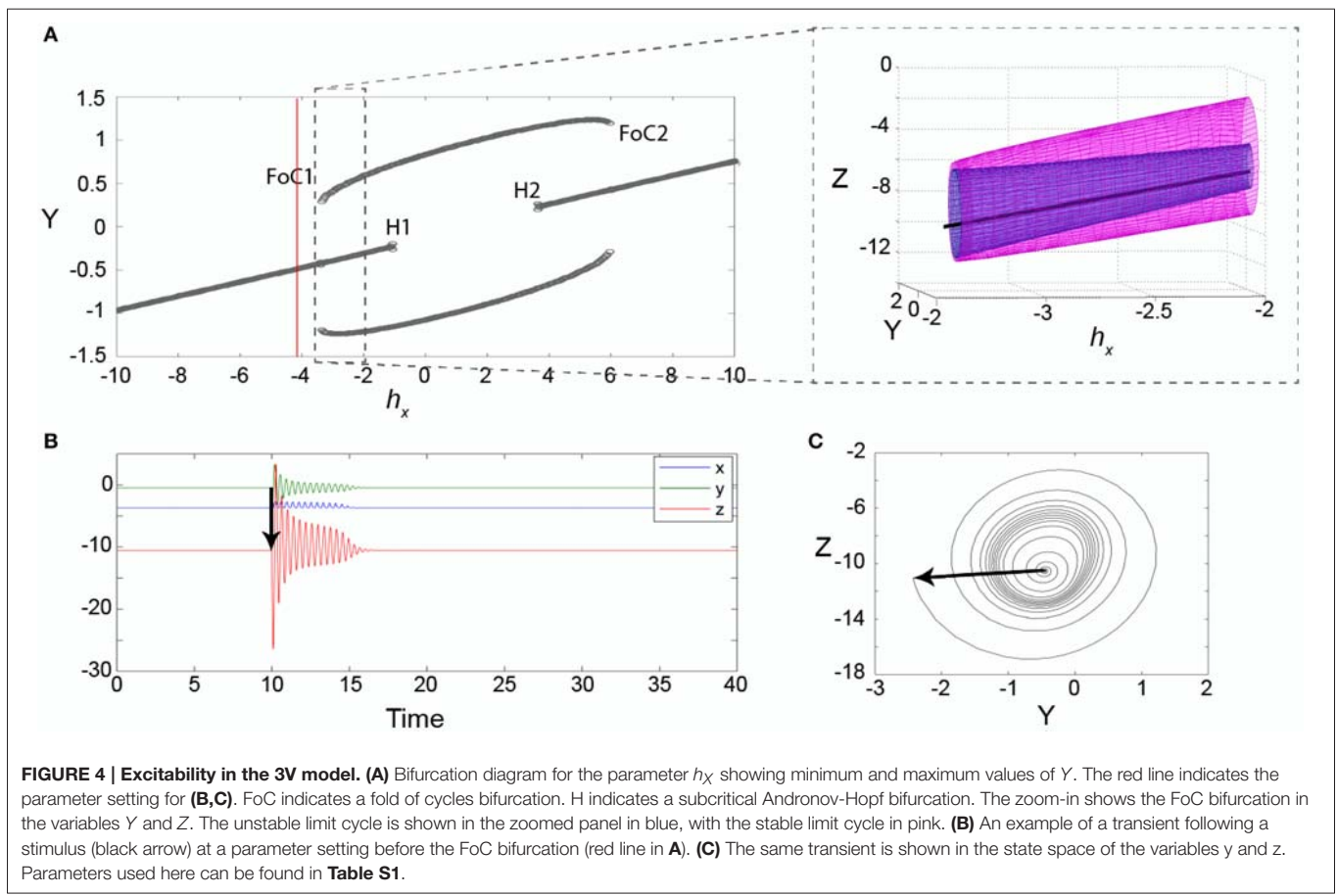

FIGURE 4 | Excitability in the 3V model. (A) Bifurcation diagram for the parameter $h_X$ showing minimum and maximum values of $Y$. The red line indicates the parameter setting for (B,C). FoC indicates a fold of cycles bifurcation. H indicates a subcritical Andronov-Hopf bifurcation. The zoom-in shows the FoC bifurcation in the variables $Y$ and $Z$. The unstable limit cycle is shown in the zoomed panel in blue, with the stable limit cycle in pink. (B) An example of a transient following a stimulus (black arrow) at a parameter setting before the FoC bifurcation (red line in A). (C) The same transient is shown in the state space of the variables y and z. Parameters used here can be found in Table S1.

## 3.2. Analysis of the Full Model

We now turn our attention to the full thalamo-cortical system. We can essentially identify the slow subsystem (*TC* and *RE*) with the slow subsystem in the 3V model ($Y$ and $Z$). The cortical system *PY* and *IN* can be considered the expansion of the 3V system variable $X$. This identification is possible through the analysis of the geometry of the fast and slow manifolds in state space (compare **Figure 3** with **Figure 6**). Where the fast manifold in the 3V model was a stable focus, this is exactly preserved in the full system (by design of the 3V system, see Section 2). Where the slow manifold in the 3V model was previously a simple sigmoid manifold, it is now a manifold that partly resembles a sigmoid manifold, only with a Hopf bifurcation at one end creating an additional limit cycle manifold in the fast subsystem. Note again, the bistable limit cycle is created around the non-linearity (or bend) of the slow manifold—as in the simplified model.

Using the knowledge from the three-dimensional reduced model, the analysis of the four-dimensional system can proceed in an analogous manner. **Figures 6C–F** show the state space projections of the full model, which can be compared to **Figures 3A–D** qualitatively. The fast manifold is still a stable focus, shown as an orange line in the projection of **Figures 6C,D**. The fast subsystem is now richer in dynamics. With respect to the slow modulation from the thalamic subsystem, the fast cortical system can undergo a supercritical Andronov-Hopf bifurcation. This means that the slow manifold is more complex in structure. **Figures 6C–F** show the slow manifold as a blue mesh. Particularly in **Figures 6C,E,F** the cone structure of the slow manifold is visible, which occurs due to the supercritical Andronov-Hopf bifurcation in the fast subsystem.

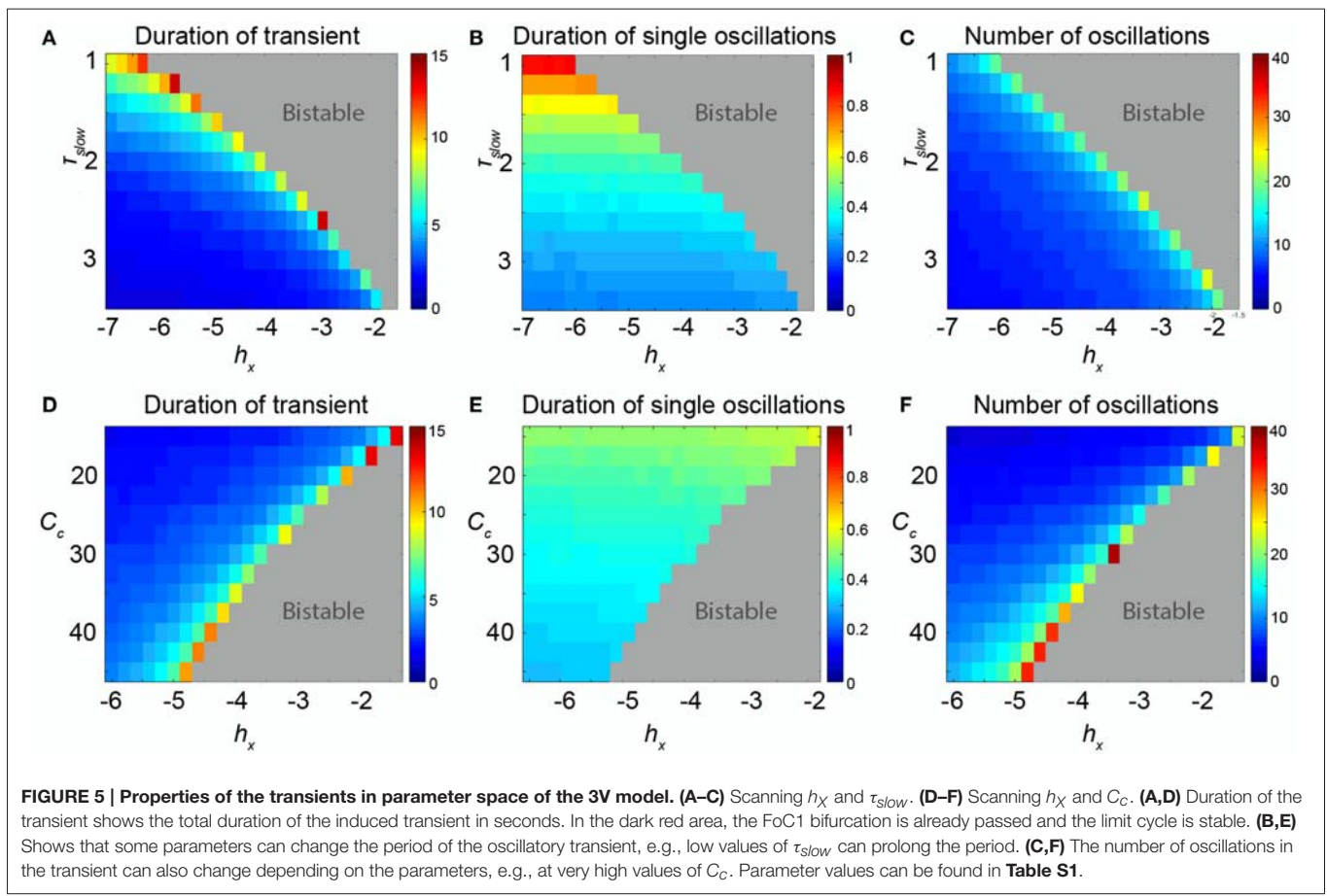

**FIGURE 5 | Properties of the transients in parameter space of the 3V model. (A–C)** Scanning $h_X$ and $\tau_{slow}$. **(D–F)** Scanning $h_X$ and $C_C$. **(A,D)** Duration of the transient shows the total duration of the induced transient in seconds. In the dark red area, the FoC1 bifurcation is already passed and the limit cycle is stable. **(B,E)** Shows that some parameters can change the period of the oscillatory transient, e.g., low values of $\tau_{slow}$ can prolong the period. **(C,F)** The number of oscillations in the transient can also change depending on the parameters, e.g., at very high values of $C_C$. Parameter values can be found in **Table S1**.

The intersection of the fast and slow manifold again forms a stable fixed point. The interaction of the slow and fast subsystem additionally creates a limit cycle. The limit cycle is not only a simple slow oscillation as in the reduced system, but also includes an additional fast spike. The fast spike arises due to the Andronov-Hopf oscillation in the fast subsystem. The attractor (black line in **Figures 6C–F**) can be characterized as having a spike-and-wave morphology (**Figure 6A**). The basin of attraction for the fixed point is now more complex than in the three-dimensional case, and more difficult to visualize as it is four-dimensional. We have illustrated it in **Figures 6C–F** as a dot cloud, and also as **Video S1**.

The bistability between the fixed point and the spike-wave limit cycle is additionally demonstrated in **Figures 6A,B**. The red and blue bars indicate two perturbations, which induce a transition to and from the spike-wave limit cycle.

The mechanism by which the bistable spike-wave limit cycle arises is, as in the three-dimensional case, through a fold of cycles bifurcation. The bifurcation scan of the system with respect to $h_{tc}$ and $h_{re}$ are shown in **Figures 7A,B**. Analogous to the 3V case, just prior to the FoC bifurcation, the transient length increases (**Figures 7C,D**). **Figure 7E** shows an example time series of such a transient. As the impending bifurcation will create a limit cycle of spike-wave morphology, the transients also show a spike-wave oscillation. **Figure 7F** shows the same time series in state space, again showing the dwelling in the region where the stable and saddle limit cycles will appear.

Desroches et al. (2012) and other studies have shown that in slow-fast systems of bursting, adjusting the time scale ratio can increase the winding numbers of the spikes in a cycle of spike and wave. In our system this is indeed the case as well. By adjusting the time scale ratio we can produce a poly-spike-wave limit cycle that is bistable to the fixed point (**Figures 8A,C**). Prior to the FoC, we can also observe transient poly-spike-wave oscillations (**Figures 8B,D**).

### 3.3. Implication for Clinical Application

As an outlook, we present a prediction for the clinical application in afterdischarges. When testing for afterdischarges clinically, variability in the stimulation response duration is often observed (e.g., in single pulse stimulation such as in Valentín et al., 2002), despite using the same stimulation parameters. In **Figure 9A** we show that in our 3V model of excitable transients simulated with low level noise input, similarly variable stimulation responses can be observed. It is important to note that the noise is of such a low level that on its own, it does not provoke high amplitude oscillatory transients in our total simulation time of 1,500 s. We stimulate such a system every 15 s, and indeed see that the response is not the same every time. Rather, short and long transients

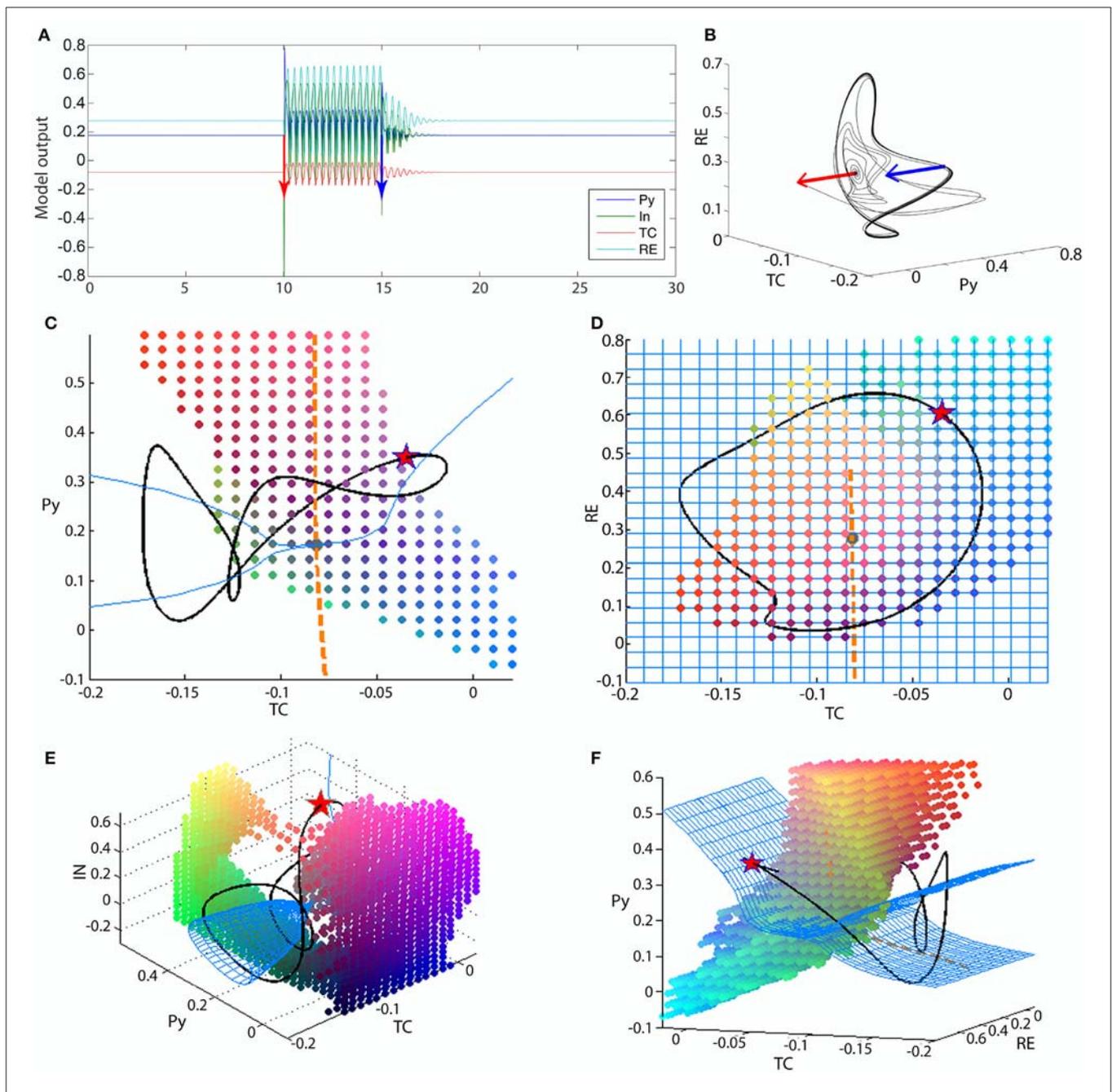

FIGURE 6 | **Bistability in the 4V model. (A)** Time series of the full thalamo-cortical system when stimulated to transition from the fixed point to the spike-wave limit cycle, and back. **(B)** PY-RE-TC state space showing the trajectory in **(A)**. Blue and red arrows indicate the stimulation points. **(C–F)** State space projection of the four-dimensional thalamo-cortical system. Blue mesh shows the slow manifold (attractors of the fast subsystem), and the orange dashed line shows the fast manifold (attractors of the slow subsystem). The intersection of the blue mesh and orange line is the fixed point for the full system (gray dot). The black line shows the spike-wave limit cycle of the full system. The colored dots outline the state space area that is the basin of attraction for the fixed point of the full system. As this basin is a 4D structure, we show a 3D slices of it here, and the slice point is a point on the SWD attractor, marked by the red star. Parameters used here can be found in **Table S2**.

are seen, despite using the same stimulation and system parameters.

To understand if there is any regularity to when shorter or longer transients are provoked, we project the simulated time series into state space in **Figure 9B** (note this is essentially the same as **Figure 4C**, only simulated with noise and zoomed in). We find the majority of the time series around the background state (fixed point) in state space. The stimuli every 15 s are seen as large displacements in the stimulation direction. We assumed the stimulation direction, i.e., to what extent each variable is affected by the stimulus, to be constant. The transient oscillatory responses are then seen as trajectories around the background state. To determine from where in state space long (vs. short) transients are provoked, we marked the stimulation position with dots in **Figure 9B**. The color of the dots indicates the provoked response length. With such a projection, we indeed see that the longer stimulation responses tend to be clustered, and located on the right hand side of the figure.

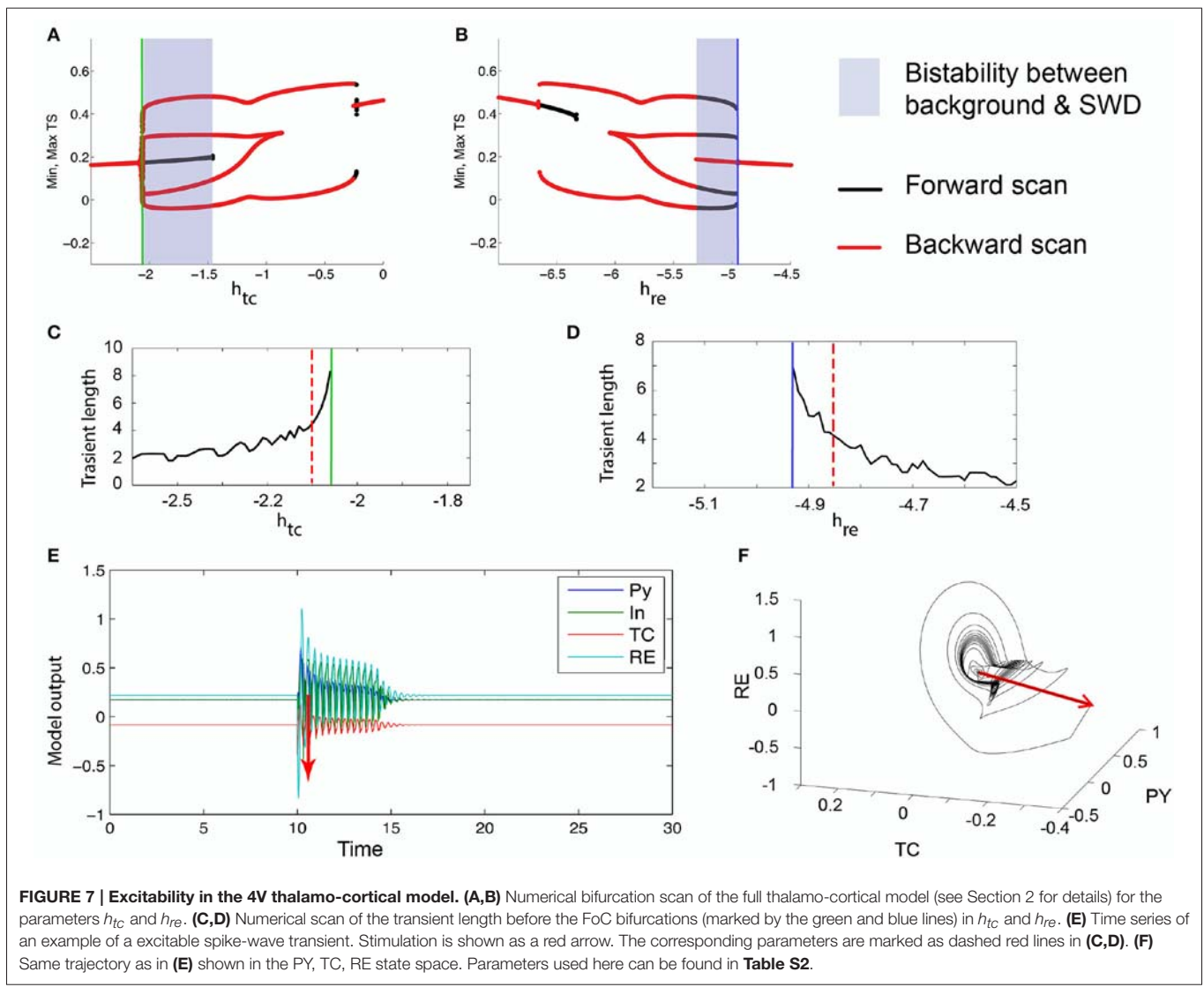

FIGURE 7 | Excitability in the 4V thalamo-cortical model. (A,B) Numerical bifurcation scan of the full thalamo-cortical model (see Section 2 for details) for the parameters $h_{tc}$ and $h_{re}$. (C,D) Numerical scan of the transient length before the FoC bifurcations (marked by the green and blue lines) in $h_{tc}$ and $h_{re}$. (E) Time series of an example of a excitable spike-wave transient. Stimulation is shown as a red arrow. The corresponding parameters are marked as dashed red lines in (C,D). (F) Same trajectory as in (E) shown in the PY, TC, RE state space. Parameters used here can be found in Table S2.

With our knowledge of the deterministic 3V system described earlier, this is easy to understand: the stimuli need to be from a position in state space that is near the ghost of the basin boundary of the background state. The stimulation direction and strength then determine where the long oscillatory transients can be provoked from. The remaining degree of variability can finally be attributed to the noisy nature of the simulation. In terms of clinical prediction, our results indicate that if we projected clinical EEG into a state space equivalent (e.g., using Taken's reconstruction Takens, 1981; Taylor et al., 2014), we might be able to observe a similar pattern, where a particular area of state space is producing long afterdischarges. The implication of this is that, whilst clinically one may simply see a varied response, a phase space reconstruction of the response may explain *why* there is a varied response. An alternative interpretation of the results is: if the background state has a dominant oscillatory component, then we predict that afterdischarges might be more readily found in a particular phase of this ongoing oscillation. This finding also holds for the full system (**Figure S1**).

## 4. DISCUSSION

We demonstrated that excitable transients can be provoked near a fold of cycles bifurcation in a reduced thalamocortical model. We also showed that this helps to explain the occurrence of excitable complex transients in the full thalamocortical model. The thalamocortical model thereby generates time series that resemble the different waveform morphologies of transient dynamics provoked by stimulation in patients (Blume et al., 2004). Hence we hypothesize that excitability in the vicinity of a FoC bifurcation is a potential dynamic mechanism underlying the clinically observed abnormal transients.

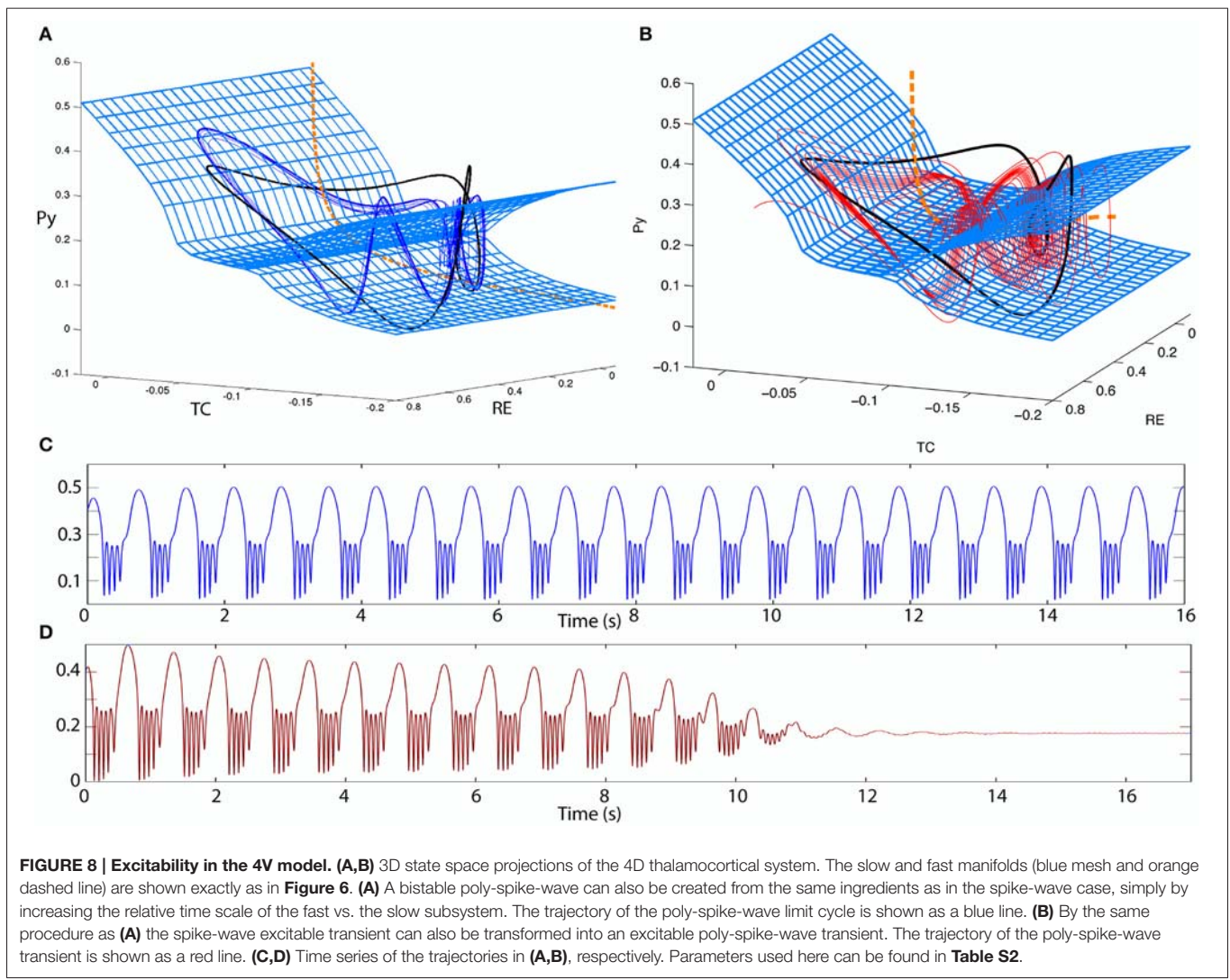

**FIGURE 8 | Excitability in the 4V model. (A,B)** 3D state space projections of the 4D thalamocortical system. The slow and fast manifolds (blue mesh and orange dashed line) are shown exactly as in **Figure 6**. **(A)** A bistable poly-spike-wave can also be created from the same ingredients as in the spike-wave case, simply by increasing the relative time scale of the fast vs. the slow subsystem. The trajectory of the poly-spike-wave limit cycle is shown as a blue line. **(B)** By the same procedure as **(A)** the spike-wave excitable transient can also be transformed into an excitable poly-spike-wave transient. The trajectory of the poly-spike-wave transient is shown as a red line. **(C,D)** Time series of the trajectories in **(A,B)**, respectively. Parameters used here can be found in **Table S2**.

## 4.1. Mechanism of Excitable Transients

The study of neural dynamics has traditionally focused on the analysis of stable states, either fixed points or limit cycles. Little effort has been devoted to the analysis of transient dynamics, with the exception of a few studies (Nowacki et al., 2011, 2012; Goodfellow et al., 2012a; Osinga and Tsaneva-Atanasova, 2013), although they were successful to explain other biological phenomena (e.g., gene expression regulation Süel et al., 2006). Traditionally, excitability in neural systems is divided in two classes: type I, and type II excitability. The distinction is made based on how the frequency of the oscillation following stimulation changes with the strength of stimulation. This has classically also been mapped to different bifurcations, near which the excitable transients are found. Type I excitability is found near saddle-node on invariant circle (SNIC) bifurcations, and type II excitability is found near an Andronov-Hopf and a fold of cycles bifurcation (Izhikevich, 2000; Buri et al., 2005). Our mechanism of excitability is very similar to the type II fold of cycles bifurcation, however the configurations of the surrounding state space are such that the dwelling behavior as described in Results is much more prominent than in other system. The dwelling behavior is mainly determined by the phase space flow, which in turn is controlled by connectivity and time scale parameters (see **Figure 5**).

In this type II of excitability, there is a soft threshold (unlike in type I excitability, where it is a hard threshold, which is the stable manifold of the saddle). The soft threshold in our case is essentially the state space area, where the stable manifold of the saddle cycle (gray tube in **Figure 3**) will form. Perturbation beyond this area will be followed by an excitable transient, which settles in an oscillation with almost stable frequency for a substantial amount of time before returning to the stable fixed point. The impending bifurcation essentially introduces a local deformation in state space that allows trajectories to dwell in the "ghost" of the to-be-born cycles (**Figure 4**). The frequency and amplitude of the excitable transient is essentially determined by the cycle that will form. Of course, this directly depends on the parameter setting of the model (**Figure 5**). The absolute duration of the excitable transient depends on the proximity to the bifurcation (as expected, as the local state space deformation nearer the bifurcation is stronger). If perturbed beyond the soft threshold the stimulus intensity will not significantly impact the morphology of the rhythmic transient, except for the initial return to the dwelling area in state space.

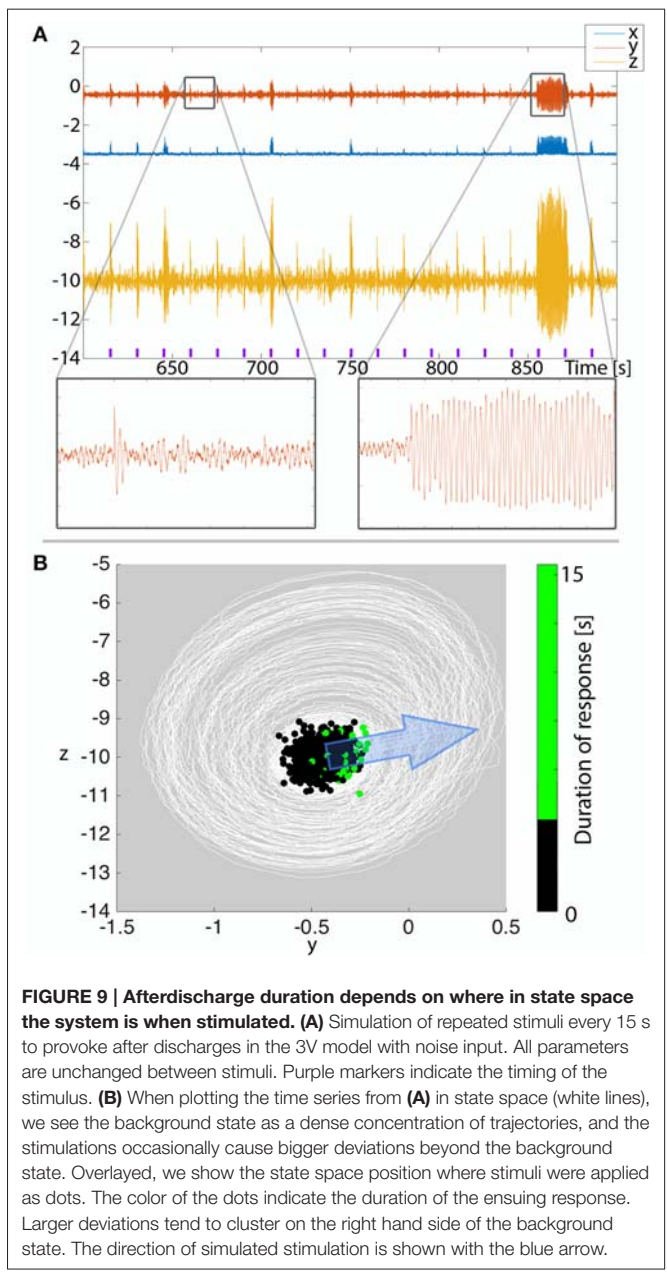

FIGURE 9 | **Afterdischarge duration depends on where in state space the system is when stimulated. (A)** Simulation of repeated stimuli every 15 s to provoke after discharges in the 3V model with noise input. All parameters are unchanged between stimuli. Purple markers indicate the timing of the stimulus. **(B)** When plotting the time series from **(A)** in state space (white lines), we see the background state as a dense concentration of trajectories, and the stimulations occasionally cause bigger deviations beyond the background state. Overlayed, we show the state space position where stimuli were applied as dots. The color of the dots indicate the duration of the ensuing response. Larger deviations tend to cluster on the right hand side of the background state. The direction of simulated stimulation is shown with the blue arrow.

These properties make the excitable transients in our model an attractive candidate for modeling stimulus induced seizure-like events (SLE). As these SLE are often referred to as complex discharges following stimulation, they are usually of a substantial duration (i.e., not a simple return to the background state). We are not aware of consistent reports of the dependency of the response amplitude or frequency on stimulus intensity. However, usually a threshold phenomenon is reported in the stimulus intensity (Lesser et al., 1984). All these properties are fulfilled in our rhythmic transient, but not in the traditional type I and type II transients (Izhikevich, 2000; Buri et al., 2005). To test if the state space setup in our model is a good candidate, a similar approach as in Jia and Gu (2012) could be used, where the system is driven by noise input. The statistics of noise induced transients (interval distributions) could hint at the type of excitability and state space setup in the data, which can be compared to real experiments, where a noisy input is provided to brain tissue.

We also highlight here, that our mechanism of creating transient oscillatory behavior is different to that of e.g., Jirsa et al. (2014). Our slow timescale does not simply serve as the slow variable change (i.e., our slow time scale is not the ultraslow time scale in variable z in Jirsa et al., 2014). Indeed, we show with our model that the ultraslow time scale/slow parameter change is not needed to create transient oscillatory behavior. This point may also have important implications for the biophysics of transient epileptiform afterdischarges, where so far the hypothesized mechanism of creating transient behavior is through such ultra-slow variables. Our model proposes that such ultra-slow variables—often hypothesized to be quantities such as extracellular potassium, or $pO_2$ Jirsa et al. (2014)—may not be needed. In this context, it is of course important to acknowledge that ultraslow time scales may still play a crucial role in the termination of seizure events or even stimulation triggered epileptiform events. For example, one compelling piece of evidence in the context of seizures is provided by Bauer et al. (2017).

Finally, we also show that the waveform morphology can be modified to reflect the various waveforms that are observed in clinical settings (Blume et al., 2004). In **Figure 7** we showed that, in addition to slow oscillations, the excitable transient can take the form of a train of spike-waves, similar to the clinical picture in **Figure 1**. This is possible through the addition of another fast variable, and the required dynamic mechanism is the introduction of a supercritical Andronov-Hopf bifurcation in the fast system during each cycle of the slow oscillation (**Figure 6**). The global bifurcation mechanism of generating the excitable transient in the first place remains the same as in the 3V system, i.e., through the fold of cycles. By adjusting the time scale separation of the two time scales, additional spikes can be added between successive waves, leading to the appearance of poly-spike-waves, also observed clinically. This spike-adding mechanism through the time-scale ratio has been well-documented in previous studies of slow-fast systems (Desroches et al., 2012).

## 4.2. Stimulation Provoked Events as Transients

In a more general context, the question arises whether excitable transients are a good model of stimulus induced seizure-like activity at all. Particularly in the context of clinical seizures, the established concept is that seizures are oscillatory attractors that a neural system can reach either through some underlying parameter change (bifurcation) (Wendling et al., 2002; Breakspear et al., 2006) or through perturbation from a coexisting stable background state (bistability) (Lopes da Silva et al., 2003; Lytton, 2008).

However, as we pointed out in the introduction, the bifurcation-based explanation requires the change of an underlying parameter for the start and termination of the seizure. This is often illustrated as a "path through parameter space" (Wendling et al., 2002; Breakspear et al., 2006; Nevado-Holgado et al., 2012). The particular attraction of the parameter variation hypothesis is that seizures (and also a small percentage of afterdischarges for that matter Blume et al., 2004) can evolve in their waveform morphology over seconds, which would point to a slow change in parameter. However, such a path has never been confirmed through direct measurements in a clinical context. Some indirect evidence exists in animal models and human EEG (Kramer et al., 2012; Jirsa et al., 2014). So, although this concept can be useful in some applications, it still needs further consolidation in order to provide a mechanistic explanation of seizure onset/offset. In the context of stimulation induced activity, such as that shown in **Figure 1**, this concept becomes particularly difficult. For the onset, the parameter change would have to take effect instantaneously triggered by the stimulus, and the termination would involve a parameter change that is not stimulation triggered. Although conceivable in theory, it is difficult to identify processes that could be measured and manipulated directly to test this hypothesis. In terms of biophysical interpretation, we essentially suggest that a slow parameter change (e.g., extracellular potassium, or $pO_2$) may not be necessary to produce stimulation induced transient epileptiform activity.

The bistability-based explanation has the advantage that stimuli can indeed induce a transition without the need for underlying parameter changes. However, for the termination, the bistable model still required a terminating stimulus, or a parameter change. Hence, excitable transients appear well suited as a deterministic model to describe stimulation induced seizure like activity. However, in a noise-driven system, the picture is more complicated. If we assume that local brain activity is modeled by noise-driven dynamics (see e.g., Breakspear et al., 2006; Deco et al., 2009; Taylor et al., 2014), even a bistable system could terminate the transient activity just through the noise input. However, this will depend on the noise level and we suggest that again, a detailed analysis of the event statistics (e.g., distribution of duration of after discharges) might offer insights into the exact dynamic mechanism. One study uses this approach and presents some indirect evidence that seizure onset may be described by a random walk process, but not seizure offset (Suffczynski et al., 2006). This is particularly interesting, as they describe seizure offset to have a deterministic component, which would fit with the excitability model.

For low noise levels (low meaning transitions induced just by the noise alone is extremely unlikely), the distinction of bistable vs. excitable would still make sense. It is easy to see that for low noise levels, the duration of afterdischarges will be much longer in the bistable case, compared to that of the excitable case. For medium levels of noise, more complex and non-trivial phenomena can arise (Lindner et al., 2004), which can be studied in future work, and matched to experimental finding (Lesser et al., 2008). In this context, it has for example been noted that an afterdischarge increases the likelihood of the following afterdischarge in the same stimulation location within a certain time window (Lee et al., 2010). Such a phenomenon is for example imaginable for an intermediate noise level driving the fast system, which can stabilize a state space region around the fixed point (see e.g., Muratov et al., 2005), enabling an increased likelihood for a subsequent excitable transient.

In this context, we also offer a prediction for the observation and interpretation of afterdischarges. Assuming that afterdischarges are indeed excitable transients, then in the presence of low level noise, the response to stimulation is variable (as observed in the clinical setting). We predict that part of this variability may be explained by the state of the ongoing background activity (**Figure 9**). If our prediction proves to be true, it does not only support our theory, but also has wide-ranging implications. For the evaluation of cortical stimulation and afterdischarges, it means that these need to be considered carefully with regards to the ongoing activity (e.g., phase of the ongoing background oscillations), and ideally over several trials. This may indeed be at the heart of the discrepancy that afterdischarges do not seem to overlap well with seizure onset zone (Blume et al., 2004).

Finally, this leads us to comment on the relationship of the stimulation induced seizure-like events, and naturally occurring seizure activity. To our knowledge, there is no direct link between the two phenomena (Blume et al., 2004), although they may be related on a fundamental level (Penfield and Jasper, 1954; Kovac et al., 2016). From the modeling perspective, the generation of excitable transients through stimulation reflects a general level of propensity/ability of the human cortex to generate seizures (as the system is near a bifurcation to seizure-like activity). However, the dynamic mechanism of seizure onset could be entirely different (as discussed, specific parameter changes could drive onset and offset, or a bistability could occur). This might explain that although afterdischarges and seizures appear to have some commonalities, the spatial sites and networks involved do not necessarily overlap. In other words, afterdischarges might be a general (not necessarily spatial) indicator of proximity to the seizure attractor, but the actual seizure onset and evolution could (through active parameter changes) involve sites that differ from those activated by local stimulation. A careful evaluation of stimulation and afterdischarges over several repeated trials might unlock the true mechanism underlying afterdischarges, and link it mechanistically to seizures.

## AUTHOR CONTRIBUTIONS

# ACKNOWLEDGMENTS

PT was funded by Wellcome Trust (105617/Z/14/Z). The funders played no role in the design or interpretation of this study.

# SUPPLEMENTARY MATERIAL

The Supplementary Material for this article can be found online at: http://journal.frontiersin.org/article/10.3389/fncom.2017.00025/full#supplementary-material

**Figure S1 | Afterdischarge duration depends on where in state space the system is when stimulated. (A)** Simulation of repeated stimuli every 10 s to provoke afterdischarges in the 4V model with noise input. All parameters are unchanged between stimuli. **(B)** When plotting the time series from **(A)** in state space (white lines), we see the background state as a dense concentration of trajectories, and the stimulations occasionally cause bigger deviations beyond the background state. Overlayed, we show the state space position where stimuli were applied as dots. The color of the dots indicate the duration of the ensuing response. Larger deviations tend to cluster on the right hand side of the background state. The direction of simulated stimulation is shown with the blue arrow.

**Table S1 |** Parameter values used to produce the figures for the reduced 3D system in this manuscript.

**Table S2 |** Parameter values used to produce the figures for the full thalamo-cortical system in this manuscript.

**Video S1 | Basin of attraction for the full thalamo-cortical system.** The basin of attraction for the full model is a 4D object. In each frame of the video, we project the basin into 3D state space of PY, IN, TC (into PY TC RE for the inset video). The slice point through the 4D space for each frame is shown as a red dot, and the video shows the full basin, where the fourth dimension is projected onto time.